\documentclass{appolb}
\usepackage{epsfig}

\def\Pom{{\bf I\!P}}

\def\lsim{\mathrel{\rlap{\lower4pt\hbox{\hskip1pt$\sim$}}
    \raise1pt\hbox{$<$}}}         

\def\gsim{\mathrel{\rlap{\lower4pt\hbox{\hskip1pt$\sim$}}
    \raise1pt\hbox{$>$}}}         

\newcommand{\be}{\begin{equation}}
\newcommand{\ee}{\end{equation}}
\newcommand{\bea}{\begin{eqnarray}}
 
\newcommand{\eea}{\end{eqnarray}}

\newcommand{\ba}{\begin{array}}
\newcommand{\ea}{\end{array}}

\newcommand{\bDelta}{\mbox{\boldmath $\Delta$}}

\newcommand{\br}{{\bf r}}

\begin{document}
\pagestyle{plain}
\newcount\eLiNe\eLiNe=\inputlineno\advance\eLiNe by -1
\title{ DIFFRACTIVE VECTOR MESONS IN DIS: \\
MESON STRUCTURE AND QCD 
\thanks{This work was partly supported by INTAS grant 97-30494}%
}
\author{Nikolai N. Nikolaev 
\address{Institut f. Kernphysik, Forschungszentrum J\"ulich,
D-52425 J\"ulich, Germany\\
and \\
L.D.Landau Institute for Theoretical Physics, Moscow, Russia}}
\maketitle

\begin{abstract}
I review the modern status of QCD theory of diffractive vector meson
production with the focus on shrinkage of photons with $Q^2$ 
and $(Q^2+m_V^2)$ scaling, 
$j$-plane properties of the QCD pomeron and Regge
shrinkage of diffraction cone, $s$-channel helicity non-conservation
and sensitivity to spin-orbital properties of vector mesons.
\end{abstract}

\section{Introduction}

There are good reasons for special interest in diffractive vector
meson production. Recall the  fundamental relationship between
the inclusive 
DIS structure function and the forward amplitude of a diagonal, 
$Q^2_f = Q_{in}^2 =Q^2$ virtual
Compton scattering (CS)
\be
\gamma^{*}_{\mu}(Q^2_{in})p\to \gamma^{*}_{\nu}(Q_f^2)p',
\label{eq:1.1}
\ee 
which for purely kinematical reasons of vanishing $(\gamma^*,\gamma^*)$
momentum transfer is diagonal in the photon helicities, ${\nu}={\mu}$.
By analytic continuation to $Q_{f}^{2}=0$ one obtains DVCS, the 
still further continuation to $Q_{f}^{2}=-m_{V}^2$ one 
obtains from CS the diffractive vector meson (VM)
production
\be
\gamma^{*}_{\mu}(Q^2)p\to \gamma^{*}V_{\nu}(\bDelta)p'(-\bDelta)\, ,
\label{eq:1.2}
\ee 
which is accessible experimentally at finite  $(\gamma^*.V)$ momentum
transfer $\bDelta$. Furthermore, the decays of VM's are
self-analyzing and azimuthal correlations of $(e,e')$, $(p,p')$ and
decay planes and polar decay angle distributions allow to reconstruct
the full set of helicity amplitudes $A_{\nu\mu}$, which allows to probe the 
mechanism of generalized CS in full complexity. The new numerical
results reported here were obtained in collaboration with Igor' Ivanov
\cite{Igor}

\section{Color dipole factorization, shrinking photons 
and $(Q^{2}+m_{V}^2)$ scaling}

The small-$x$ CS is best described in color dipole (CD) factorization, in
which  
$
A_{\nu\mu}=\Psi^{*}_{\nu,\lambda\bar{\lambda}}\otimes A_{q\bar{q}}\otimes
\Psi_{\mu,\lambda\bar{\lambda}}
$
where $\lambda,\bar{\lambda}$ stands for $q,\bar{q}$ helicities,
$\Psi_{\mu,\lambda\bar{\lambda}}$ is the wave function (WF) of the $q\bar{q}$ 
Fock state of the photon. The QCD pomeron exchange $q\bar{q}$-proton 
scattering kernel $A_{q\bar{q}}$, proportional to color dipole cross section,
does not depend on, and conserves exactly, the $q,\bar{q}$ helicities. 
For small dipoles, the CD cross section can be related to the gluon SF 
of the target, 
\be
\sigma(x,\br)\approx {\pi^2 \over 3}r^2 \alpha_{S}({A\over r^2})
G(x,{A\over r^2})\, , 
\label{eq:2.1}
\ee
where $A\approx 10$ follows from properties of of Bessel functions 
\cite{NZglue}. 
\vspace{-.3cm}

\begin{figure}[!htb]
   \centering
   \epsfig{file=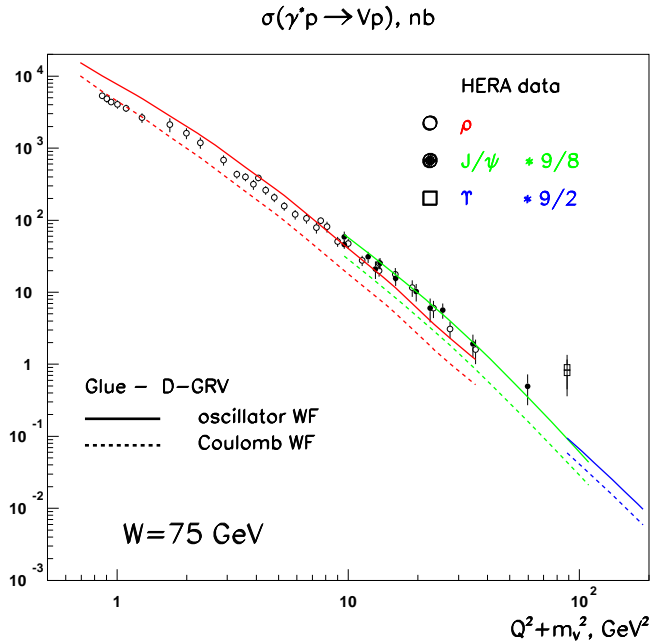,width=10cm}
\vspace{-0.5cm}
\caption{The test of the $(Q^2+m_V^2)$ scaling. The divergence of the 
solid and dashed curves indicates the sensitivity to the WF of the VM.
The experimental data are from HERA \cite{HERAdata,Bruncko}.} 
\end{figure}

The diagonal CS, i.e., inclusive DIS, probes CD cross section in broad
range of ${1 \over AQ^2} \lsim r^2 \lsim 1 $ fm$^2$ \cite{NZHERA}.
The far reaching change from diagonal CS to exclusive VM 
production is that in the final
state one swaps the pointlike photon the $q\bar{q}$ WF of which is
singular at $r\to 0$ \cite{NZ91} for the finite-size  
VM with WF which is smooth at $r\to 0$. The crucial change \cite{NN92,NNZscan} 
is that diffractive VM 
production probes the CD cross section and the VM WF at a 
scanning radius
\be
r\sim r_{S}= {6 \over \sqrt{Q^2 + m_{V}^2}}\, ,
\label{eq:2.2}
\ee
which is a manifestation of a shrinkage of the photon with $Q^{2}$. 

The three fundamental consequences of (\ref{eq:2.1}) and (\ref{eq:2.2}) are:
i) the VM production probes \cite{NNZscan} 
the gluon SF of the target at the hard scale
$\overline{Q}^2 \approx$ (0.1-0.25)$* (Q^2 + m_{V}^2)$, 
with slight variations from light
to heavy VM's, and $x=0.5(Q^2+m_{V}^2)/(Q^2+W^2)$,
ii) after factoring out the charge-isospin factors all 
VM production cross section follow a universal function of $\overline{Q}^2$,
i.e. there is $(Q^2 + m_{V}^2)$ scaling \cite{NNZscan}, see fig.~1, the
same scaling holds also
for the effective intercept $\alpha_{\Pom}(0)-1$ of the energy
dependence  of production amplitude, see fig.~2,
iii) the contribution to the diffraction slope $B$ from the $\gamma^* \to V$
transition vertex decreases $\propto r_{S}^2$ exhibiting again the 
$(Q^2 + m_{V}^2)$ scaling \cite{NZZslope}, see fig.~2.

\begin{figure}[!htb]
   \centering
   \epsfig{file=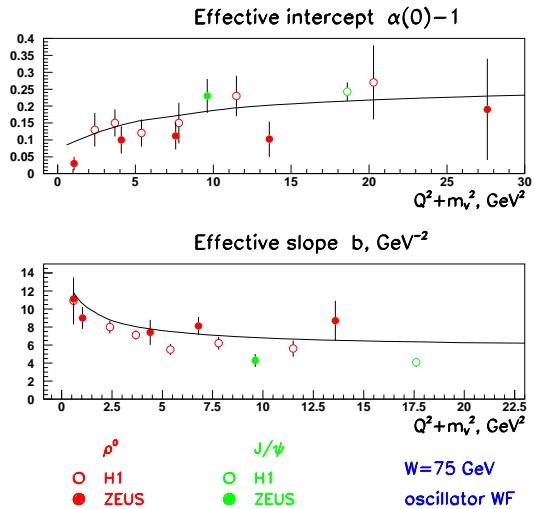,width=8cm}
\vspace{-0.5cm}
\caption{The $(Q^2+m_V^2)$ scaling of the effective intercept and diffraction slope}
\end{figure}

The agreement between theory and experiment \cite{HERAdata,Bruncko}
 is good, although there remains certain sensitivity
to not so well known WF of VM's which
can not be eliminated at the moment, see also below.

\section{Shrinkage of the diffraction cone}
 
If the pomeron is the Regge pole with the trajectory
$\alpha_{\Pom}(t)=\alpha_{\Pom}(0)+\alpha_{\Pom}'t$, then the 
diffraction slope would rise with energy,
$B(W^2)=B_{0}+ 2\alpha_{\Pom}'\log(W^{2}/W_{0}^2)$.
The common prejudice based on scaling $\alpha_S=$const approximation
is that the BFKL pomeron is a fixed branching point, 
i.e., $\alpha_{\Pom}'=0$,
with no shrinkage of the diffraction cone. Which is almost
tautological 
because in this approximation one is short of any length scale. This
toy model is nice because it is exactly solvable but it is not QCD, 
in which the asymptotic freedom (AF), i.e., running coupling $\alpha_S$,
and the fact that perturbative gluons have finite propagation
radius, introduce the perhaps related length scale. The consistent
implementation of AF into color dipole BFKL equation has been done
by Zakharov, Zoller and myself in 1994 \cite{NZZletters}. The corresponding
QCD pomeron has been proven to be a series of moving Regge poles
\cite{NZZspectrum}. As a matter of fact, already in 1975 
Fadin, Kuraev and Lipatov noticed that AF brings about the 
fundamental transformation of the QCD pomeron from a
fixed branching point to a series of moving poles \cite{FKL}.
With the specific infrared regularization
used in   \cite{NZZletters,NZHERA,NZZspectrum} we found 
$\alpha_{\Pom}'\approx 0.07$
GeV$^{-2}$for the rightmost hard BFKL 
pole and a somewhat smaller slope for trajectories of
subleading poles. Under plausible boundary condition, the 
interference of the rightmost and subleading pomerons was shown
to produce a shrinkage with $\alpha_{eff}' \sim 0.15$ GeV$^{-2}$
\cite{NZZslope}.
Such a sensitivity of shrinkage of the diffraction cone to subleading
Regge components in $pp$ and $\bar{p}p$ scattering is old news.

Our fundamental prediction of shrinkage of the diffraction 
cone for hard diffractive DIS
has been confirmed recently by the ZEUS collaboration \cite{ZEUSDIS2000}, 
which measured the energy dependence of diffraction slope
for the $J/\Psi$ photoproduction with the result
$\alpha_{\Pom}'=0.098 \pm 0.035(stat)\pm0.050(syst)$, which
is consistent with our numerical results \cite{NZZslope}.

\section{Pomeron helicity-flip and breaking of SCHC}

As emphasized above, the helicity of quarks in  $q\bar{q}$-target scattering 
is conserved exactly, which for long has been believed to entail
the $s$-channel helicity conservation (SCHC). The fundamental point
is that the sum of quark and antiquark helicities equals helicity of 
neither photon nor vector meson. Only for the 
nonrelativistic massive quarks, $m_{f}^{2} \gg Q^{2}$ the only allowed 
transition is $\gamma^{*}_{\mu} \to q_{\lambda} +\bar{q}_{\bar{\lambda}}$ 
with $\lambda +\bar{\lambda}=\mu$. In the relativistic case transitions
of transverse photons $\gamma^{*}_{\pm}$ into the $q\bar{q}$ state with 
$\lambda +\bar{\lambda}=0$, 
in which the helicity of the photon is transferred to the $q\bar{q}$ orbital 
 momentum, are equally allowed. Consequently, the QCD pomeron exchange 
SCHNC transitions
$\gamma^{*}_{\pm} \to (q\bar{q})_{\lambda +\bar{\lambda}=0} \to
\gamma^{*}_{L} ~~~{\rm and}~~~
\gamma^{*}_{\pm} \to (q\bar{q})_{\lambda +\bar{\lambda}=0 }\to
\gamma^{*}_{\mp} $ 
are allowed \cite{NZDIS97,NPZLT} and SCHNC persists at small $x$. 
We emphasize that the above argument 
for SCHNC does not require the applicability of pQCD.
Furthermore, the leading  contribution to the proton structure function
comes entirely from SCHNC transitions of transverse photons - the
fact never mentioned in textbooks.

\begin{figure}[!htb]
   \centering
   \epsfig{file=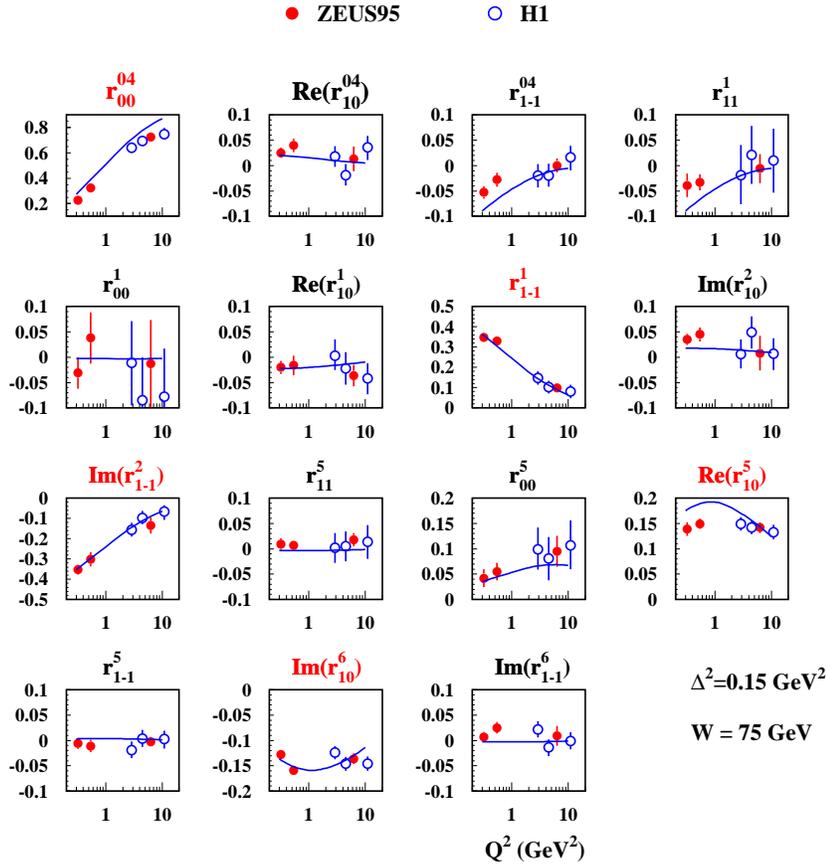,width=12cm}
\vspace{-0.3cm}
\caption{Predictions for the spin density matrix in the $\rho^{0}$
production vs. the experimental data from HERA.}
\end{figure}

The first ever direct QCD evaluation \cite{NZDIS97} of SCHNC effect - the
 LT-interference 
of transitions $\gamma^{*}_{L}p \to p'X$ and $
\gamma^{*}_{\pm}p \to p'X$ into the same continuum 
diffractive states $X$  - has been reported by Pronyaev, Zakharov and myself
in 1997.
 Experimentally, it can be measured at HERA by both H1 and
ZEUS via azimuthal correlation between the $(e,e')$ and 
$(p,p')$ scattering planes and can be used the determination of the otherwise elusive 
$R=\sigma_{L}/\sigma_{T}$ for diffractive DIS is found in \cite{NPZLT}. 
The principal issue is that this asymmetry persists, and even rises slowly, at small 
$x_{\Pom}$.

SCHNC helicity flip only is possible due to 
the transverse and/or longitudinal Fermi motion of quarks and
is extremely sensitive to spin-orbit coupling in the
vector meson, I refer for details to \cite{KNZ98,IN99}.
The consistent analysis of production of 
$S$-wave and $D$-wave vector mesons is presented only
in \cite{IN99}.  One would readily argue  based on the results 
\cite{NZDIS97,NPZLT} that by exclusive-inclusive 
duality \cite{GNZlong} between diffractive
 DIS into continuum and vector mesons the dominant 
SCHNC effect in vector meson production is the interference 
of SCHC $\gamma^{*}_{L} \to V_L$ and SCHNC $\gamma^{*}_{T}\to V_L$ production,
i.e., the element $r_{00}^{5}$ of the vector meson spin
density matrix. The overall agreement
between our theoretical estimates 
\cite{Igor} of the 
spin density matrix $r_{ik}^{n}$ for
diffractive  $\rho^{0}$ assuming pure $S$-wave in the $\rho^{0}$-meson
and the ZEUS \cite{ZEUSflip} and H1 \cite{H1flip} experimental data
is very good. There is a clear evidence for  $r_{00}^{5}\neq 0$, see
fig.~3.
\vspace{-0.3cm}
\begin{figure}[!htb]
   \centering
   \epsfig{file=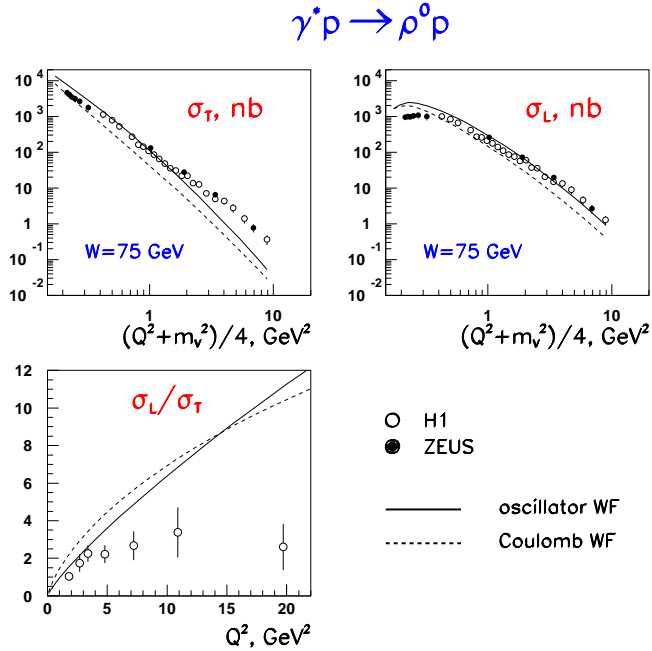,width=10cm}
\vspace{-0.3cm}
\caption{The demonstration that soft WF of the $\rho^0$ underestimates 
$\sigma_{T}$ at large $Q^2$.} \vspace{-.3cm}
\end{figure}

\section{A fly in the pie: the $\sigma_L/\sigma_T$ puzzle?}

In fig.~4 we show separately the predictions for $\sigma_L$ and $\sigma_{T}$.
Evidently, the toy models with soft wave functions for VM fail at
large $Q^{2}$. The natural interpretation is that these toy models
underestimate the admixture of small size color dipoles in vector mesons.

Indeed, consider $R_{el}= \sigma_{L}/\sigma_T$ for elastic CS 
$\gamma^*p\to \gamma^*p$, which is quadratic in the ratio of
CS amplitudes. By optical theorem one finds
\be
R_{el}= \left|{A(\gamma^*_L p\to \gamma^*_L p) \over 
A(\gamma^*_T p\to \gamma^*_T p)}\right|^2= 
\left({\sigma_{L}\over \sigma_T}\right)^2_{DIS} \approx 4\cdot 10^{-2}
\label{eq:4.1}
\ee
Here I used the prediction \cite{NZHERA} for inclusive DIS
$R_{DIS} = \left.\sigma_{L}/ \sigma_T\right|_{DIS}\approx 0.2$, which
is consistent with the indirect experimental evaluations at HERA.
The result $R_{el} \ll 1$ for diagonal CS with production of the 
pointlike final state photon must be contrasted to 
theoretical expectation $R \sim Q^2/m_V^2 >> 1$ for non-pointlike
vector meson production. Such a dramatic change from $R_{el}$ to $R$
for VM's suggests that predictions for $R$ are extremely sensitive to
admixture of quasi-pointlike $q\bar{q}$ in VM. Evidently, such an
admixture would lower the theoretical results for $R$ for the $\rho^o$
and the possible elimination of the 
observed disagreement between experiment and theoretical
evaluations of $R$ based on too crude a 
soft WF of VM's is good news!
A consistent 
theoretical analysis of the short distance WF of VM's is as yet lacking.

\section{Helicity flip and spin-orbit coupling in VM's}

In the $D$-wave state the total spin of $q\bar{q}$ pair is predominantly 
opposite to the spin of the $D$-wave vector meson. As a results, SCHNC 
in production of $D$-wave vector mesons is much stronger \cite{IN99} 
than for the ground state S-wave mesons, which 
may facilitate the long disputed $D$-wave 
vs. $2S$-wave assignment of the $\rho'(1480)$ 
and $\rho'(1700)$ and of the $\omega'(1420)$ and $\omega'(1600)$.
Striking predictions for D-wave vector meson production include 
\cite{Igor,IN99} abnormally large 
higher twist corrections \cite{IN99} and non-monotonous  $Q^2$ 
dependence of  $R^{D}=\sigma_{L}/\sigma_{T}$. 
\begin{figure}[!htb]
\vspace{-.3cm}
   \centering
   \epsfig{file=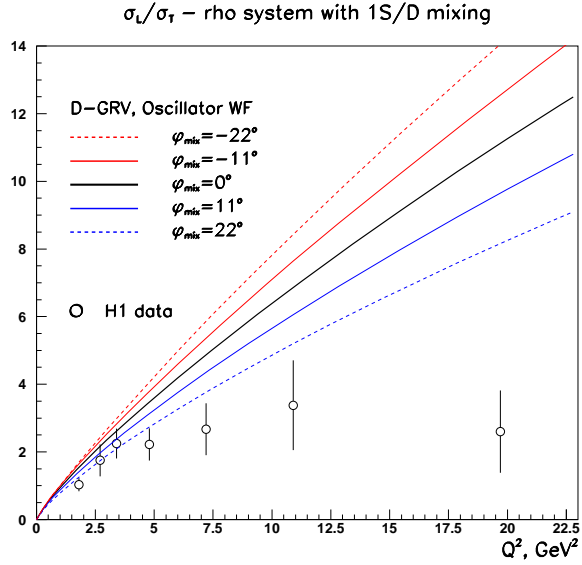,width=9cm}
\vspace{-0.2cm}
\caption{The sensitivity of $R=\sigma_L/\sigma_T$ for
$J/\Psi$ production to the S-D-mixing.}
\vspace{-.3cm}
\end{figure}

Besides that, for D-wave vector mesons we predict anomalously small
$\sigma_L/\sigma_T$ which by virtue of S-D mixing could affect $R$ for
the ground state vector mesons. 
As well known, all popular confining potentials give rise to the tensor force.
Recall a large S-D mixing angle,
$\phi_{SD} \sim 14^o$, in an even such
a loosely bound system as a deuteron. 
The only well established 
D-wave quarkonium is $\Psi(3770)$, for which the pure D-wave
assignment suggests the leptonic decay width 
$\Gamma(\Psi(D)\to e^+e^-) = 0.046\ \mbox{keV}$ to be 
contrasted to the experimental finding $
\Gamma_{exp}(\Psi(3770)\to e^+e^-) = 0.26\ \mbox{keV}
$. Attributing the enchancement of the leptonic decay width 
to the mixing with the $J/\psi(1S)$, one finds two
solutions for the S-D-mixing angle,
$
\phi_{SD} \approx 23^\circ\,,\qquad 
\phi_{SD} \approx -9^\circ\, .
$
The results presented in fig.~5 show that $R$ can be lowered 
substantially and $\sigma_L/\sigma_T$ puzzle can be eliminated 
to a large extent at the expense of admissible $S/D$ mixing.

\section{Conclusions }

\begin{itemize}
\item Consistent use of the recently determined  unique differential gluon 
structure function of the proton \cite{INdiffglue}
within $\kappa$-factorization approach
eliminates the sensitivity of vector meson production amplitudes to
the gluon structure function of the proton.
\item Consistent incorporation of $S$ and $D$ wave vector
meson spinorial structures allows for analysis of either
 pure $S$ and $D$ states or their  mixture.
\item Predictions of  $\kappa$-factorization approach
 are in agreement with experiments both on ground and excited vector mesons;
the only discrepancy --- underprediction of 
$\sigma_T$ at large $Q^2$ --- signals 
that presently used soft wave function Ans\"atze do 
not exhaust the whole physics
at short distances.
\item We predict very different behavior of basic $1S$/$2S$/$D$ state
observables.
\item A large part of $\sigma_L/\sigma_T$ puzzle can be eliminated
at the expense of strong  $S/D$ mixing in $\rho$ system;
the relatively large $e^+e^-$ decay width of $\psi(3770)$
suggests that mixing indeed can be strong.
\end{itemize}

Thanks are due to organizers of Meson-2000 for the most exciting
workshop.


\begin{thebibliography}{88}
\addcontentsline{toc}{section}{References}

\bibitem{Igor}   
I.P. Ivanov \& N.N. Nikolaev, 
Talk given at DIS 2000, Liverpool, England, 25-30 Apr 2000. 
e-Print Archive: hep-ph/0006101 

\bibitem{NZglue} 
N.N.Nikolaev and B.G.Zakharov,
{\sl Phys. Lett.} {\bf B332} (1994) 184.

\bibitem{NZHERA} 
N.N. Nikolaev, B.G. Zakharov, {\em Phys. Lett.}
 {\bf B327}  (1994) {157}.

\bibitem{NZ91} 
N.N.~Nikolaev and B.G.~Zakharov, {\em Z. Phys.} {\bf C49} (1991) 607

\bibitem{NN92} 
N.N. Nikolaev, {\sl Comments Nucl. Part. Phys.} {\bf 21} (1992) 41


\bibitem{NNZscan} 
J. Nemchik, N.N. Nikolaev and B.G. Zakharov,
{\sl Phys. Lett.} {\bf B341} (1994) 228

\bibitem{NZZslope} 
N.N. Nikolaev, B.G. Zakharov and V.R. Zoller, 
{\sl Phys. Lett.} {\bf B366} (1996) 337;
J. Nemchik, N.N. Nikolaev, E. Predazzi, B.G. Zakharov 
and V.R. Zoller, {\sl J. Exp. Theor. Phys.} {\bf 86}, 1054 (1998)

\bibitem{HERAdata} 
H1 Collab: C. Adloff et al., {\bf 
  DESY 99-010};   ZEUS Collab: J.\ Breitweg et al.): {\sl Eur.\ Phys.\
  J.} {\bf  C6}, 603 (1999) and references therein


\bibitem{Bruncko}  
D. Bruncko, talk at Meson-2000, these proceedings.




\bibitem{NZZletters} 
N.N. Nikolaev, B.G. Zakharov and V.R. Zoller, 
{\sl JETP Lett.} {\bf 59} (1994) 6

\bibitem{NZZspectrum} 
N.N. Nikolaev, B.G. Zakharov and  V.R. Zoller, {\sl JETP Lett.} {\bf 66}
(1997) 138

\bibitem{FKL} 
 V.S.Fadin, E.A.Kuraev and L.N.Lipatov  {\sl Phys. Lett.}
{\bf B60} (1975) 50;
E.A.Kuraev, L.N.Lipatov and V.S.Fadin, {\sl Sov.Phys. JETP}
{\bf 44} (1976) 443; {\bf 45} (1977) 199.

\bibitem{ZEUSDIS2000} 
ZEUS Collab: A. Bruni, Talk given at DIS 2000, Liverpool, 
England, 25-30 Apr 2000. 

\bibitem{NZDIS97} 
N.N. Nikolaev and B.G. Zakharov,
Deep Inelastic Scattering and QCD, Proceedings of DIS'97, 
Chicago, IL, 14-18 April 1997, AIP Conference Proceedings No.407, editors
J. Repond and D. Krakauer, pp. 445-455. 


\bibitem{NPZLT} 
N.N. Nikolaev, A.V. Pronyaev and B.G. Zakharov, {\sl Phys. Rev.} {\bf D59}
091501 (1999)


\bibitem{KNZ98} 
E.V. Kuraev, N.N. Nikolaev and B.G. Zakharov, 
{\sl JETP Lett.} {\bf 68} (1998) 667. 

\bibitem{IN99} 
I.P.Ivanov and N.N.Nikolaev, {\sl JETP Letters} {\bf 69} (1999) 268.

\bibitem{GNZlong} 
M.Genovese, N.Nikolaev and B.Zakharov, 
{\sl Phys. Lett.} {\bf B380},  213 (1996).

\bibitem{ZEUSflip} 
ZEUS Collab: A. Savin, DIS'99 and {\bf DESY-99-102}.

\bibitem{H1flip} 
H1 Collab: B. Clerbaux, DIS'99 and {\bf DESY-99-010}.


\bibitem{INdiffglue}
I.P. Ivanov and  N.N. Nikolaev, 
e-Print Archive: hep-ph/0004206 



\end{thebibliography}
\end{document}